\documentclass[journal]{IEEEtran}

\usepackage[caption=false]{subfig}

\usepackage[numbers]{natbib}

\usepackage{url,hyperref}
\usepackage[switch]{lineno} 

\usepackage{microtype}
\usepackage{graphicx}
\usepackage{indentfirst}
\usepackage{array,multirow}
\usepackage{fancyhdr}
\usepackage{caption}
\usepackage{float}
\usepackage{booktabs}
\usepackage{times} 
\usepackage{CJKutf8}
\usepackage{fontawesome5}
\usepackage{color}
\usepackage{comment}
\usepackage{svg}
\usepackage{here}
\usepackage{xcolor}
\usepackage{CJKutf8}
\usepackage{algorithmicx,algorithm,algpseudocode}
\usepackage{amsmath,amsfonts,bm}
\usepackage{soul}
\usepackage[normalem]{ulem}
\useunder{\uline}{\ul}{}

\newcommand{\ctext}[1]{\raise0.2ex\hbox{\textcircled{\scriptsize{#1}}}}

\newcommand{\eg}{e.\,g.,\ }
\newcommand{\ie}{i.\,e.,\ }

\begin{document}

\title {Pre-instruction for Pedestrians Interacting Autonomous Vehicles with eHMI: Effects on Their Psychology and Walking Behavior}

\author{
Hailong Liu$^{1,*,\dagger}$,~\IEEEmembership{Member, IEEE} and Takatsugu~Hirayama$^{2}$,~\IEEEmembership{Member, IEEE}

\thanks{$^{1}$~Hailong Liu is with Graduate School of Science and Technology, Nara Institute of Science and Technology, 8916-5 Takayama-cho, Ikoma, Nara, 630-0192, Japan. }
\thanks{$^{2}$~Takatsugu~Hirayama is with Faculty of Environmental Science, University of Human Environments, 6-2, Kamisanbonmatsu, Motojuku-cho, Okazaki, Aichi, 444-3505, JAPAN and Graduate School of Informatics, Nagoya University, Furo-cho, Chikusa-ku, Nagoya, Aichi, 464-8601, JAPAN.}
\thanks{$^{*}$~Corresponding Author: Hailong Liu, Email: liu.hailong@is.naist.jp}
\thanks{$^{+}$~This study was completed while the first author Hailong Liu was affiliated with Graduate School of Informatics, Nagoya University in 2019.4-2021.10.}
}
\maketitle

\begin{abstract}
External human-machine interface~(eHMI) is considered as a new explicit communication method for pedestrian-AV interactions, particularly in encounter scenarios. 
Pedestrians without prior negotiation experience with eHMI may misinterpret the driving intentions of AV, leading to confusion and unpredictable behavior.
To address this, our study suggests providing pre-instruction on eHMI to enhance comprehension. 
To compare pedestrians' subjective feelings and walking behavior changes with and without the use of eHMI, as well as before and after receiving pre-instructions, a road crossing experiment using a within-subject design was conducted.
In the experiment, the participants were challenged to recognize situations and experienced uncertainty when encountering AVs lacking eHMI, in contrast to manual driving vehicles. 
After the pre-instruction, participants could understand the driving intention of an AV with eHMI and predict its driving behavior more easily. 
Furthermore, participants' subjective feelings and hesitation to make decisions improved to align with the same criteria as encountered with a manual driving vehicle.
Additionally, this study found that the information guidance effect of using eHMI makes participants' walking speeds more consistent over multiple trials after pre-instruction.
\end{abstract}


\section{INTRODUCTION}

\IEEEPARstart{E}{nsuring} safe and reliable interactions between pedestrians and vehicles when they encounter each other depends on establishing effective communication. 
This effective communication allows both parties to understand each other's intentions and predict each other's behaviors~\citep{li2021autonomous,dey2021communicating}.

In interactions between manual driving vehicles (MV) and pedestrians, the driving intentions of the drivers of MVs are typically inferred through implicit communication cues, including vehicle speed, acceleration and driving direction~\citep{rasouli2019,tian2023deceleration}.
Furthermore, employing explicit communication cues, such as a head nod and hand gestures, assist pedestrians in comprehending the driver's intention and resolving negotiation ambiguities during interactions~\citep{sucha2017pedestrian,farber2016communication,onkhar2021towards}.
Particularly, in complex urban environments and shared spaces where explicit communication methods are frequently used between MV's drivers and pedestrians today~\citep{Vissers2016}.
For example, when a pedestrian encounters an MV on a narrow road or in a parking lot, the driver can clearly communicate their intention and quickly reach an agreement to negotiate with the pedestrian through eye contact, hand gestures, and verbal communication.

Autonomous vehicles~(AVs) with levels 3--5 automated systems are expected to become popular in the near future~\citep{10319096}.
As a result, interactions between pedestrians and AVs will become more frequent~\citep{li2021autonomous,wang2023my}.
Especially in traffic scenarios such as shared spaces, unsignalled intersections, narrow roads, and parking lots, the interactions between AVs and pedestrians are expected to increase.
Drivers (users) of levels 3-5 AVs are not necessarily responsible for driving tasks when using these ADSs in their operational design domain~\citep{SAE_j3016_2016}.
Therefore, issues may arise, such as difficulties in communicating the driving intentions of AVs to pedestrians and negotiating the right-of-way with them~\citep{li2021autonomous}.
This may make it difficult for pedestrians to understand the intentions of AVs quickly and clearly~\citep{liu2020when,Liu2022_APMV}.
Moreover, potential issues are caused, such as safety hazards, inefficiency, and poor prosociality~\citep{batson2003altruism,wang2023my}.
Thus, understanding how AVs should communicate with pedestrians has become a pressing concern.

\begin{figure*}[t]
\centering
\includegraphics[width=1\linewidth]{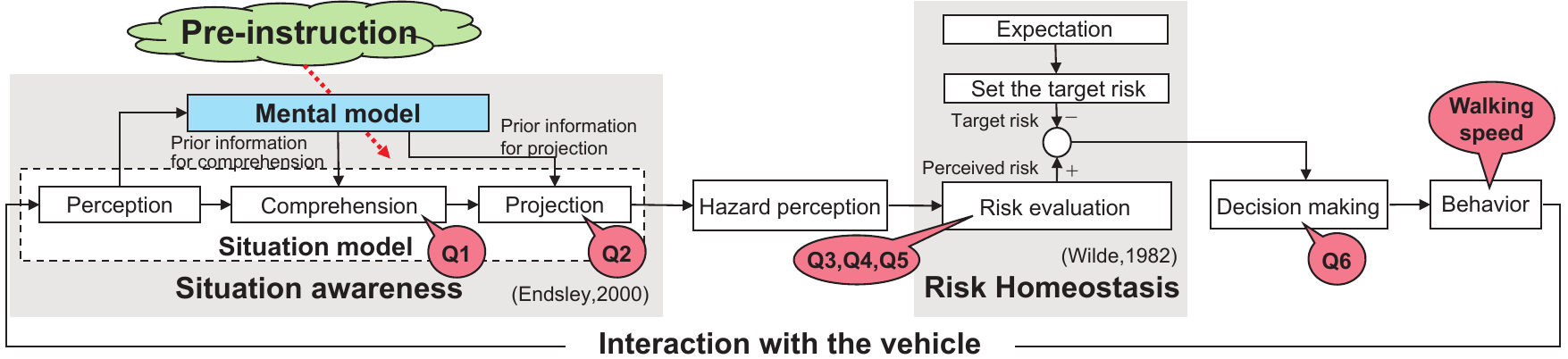}
\caption{Pre-instruction for the calibration of the mental model based on the cognition-decision-behavior model proposed in~\citep{Liu2022_APMV}. The Q1 to Q6 are subjective evaluation questions designed based on this model were used in the experiments (see section~\ref{sec:SE_Q}).}
\label{fig:model}
\end{figure*}

\subsection{External Human-Machine Interface and Its Issues}

A new communication approach using an external human-machine interface~(eHMI) is considered one of potential solutions to the issues in pedestrian-AV interactions~\citep{schieben2019designing,Clercq2019}. 
In particular, various studies evaluated the effectiveness of eHMIs in presenting the intentions of AVs to pedestrians using light bars, icons, and text~\citep{rettenmaier2019passing,Stefanie2020,faas2020external,dey2021communicating}.
Although these studies advocated for good eHMI designs, they could not ensure that pedestrians fully recognize the intentions of AVs, the rationale underlying the intentions of AVs, and the functional limitations of AVs through these eHMIs. 
In particular, pedestrians with limited experience in interacting with an AV equipped with an eHMI were more affected~\citep{Michal2020,lee2022learning,lee2024hello}.
For example, even if pedestrians clearly recognize that the AV has yielded the right of way through the message ``you go first'' displayed on the eHMI, their understanding may differ regarding when, how, or why the AV displays this information, especially when the sensor range is unknown. 
Furthermore, pedestrians may be uncertain of the time available to cross the road when the message is displayed because it is difficult to predict when the vehicle will start moving again.
If pedestrians are struggling to understand driving intentions and predict the driving behaviors of an AV, they may feel threatened and exhibit unpredictable walking behaviors.

The above-mentioned problem of eHMIs can be solved by increasing the experience of pedestrians interacting with AVs to understand the intention of the AV from the information on eHMI~\citep{Michal2020,lee2022learning}.
However, this method often requires pedestrians to engage in ongoing trial and error, imposing a substantial time investment for them to attain proficiency in interacting with AVs.
Furthermore, pedestrians may experience further risk because they do not fully understand the intention of the AV during this learning process.

Several related studies have reported that instructions for an in-vehicle human-machine interface (iHMI) of autonomous vehicles (AVs) have allowed drivers to enhance their understanding of using AVs~\citep{forster2019user,matsuo2024enhancing}, interact effectively with AVs, and increase their trust in AV~\citep{hergeth2017prior, Sonoda_SA_Trust, edelmann2020effects}. 
However, due to the growing popularity of AVs, instructing pedestrians to understand the intention of the AV correctly during interactions presents an urgent issue that has yet to be widely studied.

\subsection{Human Cognition and Knowledge Learning}

This study suggests that the mechanism of the cognition-decision-behavior process of pedestrians is important during interactions with an AV to help pedestrians correctly understand the intentions of AVs through an eHMI.
Therefore, this study focuses on the model proposed by~\citep{liu2020when,Liu2022_APMV} as presented in Fig.~\ref{fig:model}.
This model consists of three parts, \ie situation awareness, risk evaluation based on hazard perception, and decision-making for behavior generation based on risk homeostasis.

Situation awareness can be viewed as a current instantiation of the mental model~\citep{endsley2000}, \ie it is a prediction model constructed and supported by an underlying mental model~\citep{endsley1995toward,mogford1997mental}. 
Specifically, the situation awareness of a pedestrian includes the ability to perceive objects in the surrounding environment (\ie perception), understand the state and intention (\ie comprehension), and predict the future state (\ie projection). 
A mental model is a highly organized and dynamic knowledge structure that contains meaningful declarative and procedural knowledge derived from experiences~\citep{jones2011mental,Al-Diban2012}. 
It is generated spontaneously by repeatedly recognizing and interpreting the target system~\citep{staggers1993mental}. 
In summary, our perspective is that the mental model of people provides their situational model with some prior information and knowledge to guide them in understanding and predicting specific situations.
Pedestrians perceive hazards based on predicted outcomes and evaluate associated risks (\ie subjective risk).
Then, they compare the subjective risk with their acceptable risk level (i.e., target risk) and determine their walking behaviors. 
This risk compensation process can be explained by the risk homeostasis theory~\citep{wilde1982theory}.

The above process shown in Fig.~\ref{fig:model} has been verified via a causal discovery for a study of interactions between pedestrians and autonomous personal mobility vehicles~\citep{liu_IV24}.

\subsection{Purpose}\label{sec:PURPOSE}

Based on the above studies, the purpose of this study is that uses an essential approach to solving the problems addressed to ensure pedestrians quickly establish the correct mental model of an AV with an eHMI through the pre-instruction.
In particular, the approach is proposed to instruct pedestrians to understand the execution conditions, mechanisms, and functional limitations of the AV with eHMI. 

We reported a brief description of the proposed method and preliminary results on pedestrians' subjective evaluations in~\citep{liu2021importance}.
Based on this pre-study, this study not only further discusses the impact of the proposed method on the subjective evaluations, but also improves the analysis methods and further analyzes the walking behaviors of the pedestrians.
Therefore, this study aims to conduct a field experiment to verify the effects of the proposed pre-instruction on pedestrian psychology and walking behaviors.

\subsection{Hypothesis}

Based on the above arguments, this study proposes the accurate formation and calibration of a mental model of the AV with the eHMI by instructing pedestrians with relevant knowledge of the eHMI to improve their experience of interaction with the AV.
Our objective is to analyze the influence of the pre-instruction of the eHMI's rationale on the situation awareness, subjective feelings and walking behavior of pedestrians when they encounter the AV.

This study proposes the following hypotheses and validates them through an experiment:
\begin{enumerate}
\item[H1:] Pedestrians who correctly understand the rationale of eHMI through the pre-instruction exhibit improved situation awareness, subjective feelings, and decision-making during the interaction.
\item[H2:] Participants' walking behavior becomes more consistent over multiple interactions as they gain a comprehensive understanding of eHMI principles through the pre-instruction.
\end{enumerate}

\section{METHOD}\label{sec:experiment}

This study applied a Wizard of Oz experimental design to ensure the safety of the experiment. 
Using a ghost driver trick~\citep{rothenbucher2016ghost}, a driver-less AV was simulated using an MV driven by an expert driver. 
This experiment simulates a scene in a shopping center parking lot where a pedestrian wants to cross the road to use an elevator, as shown in Fig.~\ref{fig:map}.
Participants will encounter a car when crossing the road in four designed interaction scenarios.
Participants need to decide whether to cross the road.
This experiment was conducted in a blocked area of the B2F parking lot of Toyota Stadium, Toyota-shi, Aichi, Japan.
This experiment complied with the Declaration of Helsinki and it was approved by the ethics review committee of the Institute of Innovation for Future Society, Nagoya University (No.~2021-12).

\subsection{Pedestrians}
G*Power 3.1.9.2 (effect size $f=0.25$, $\alpha=0.05$, $power=0.85$) suggested that the minimum sample size was 26 for repeated measures ANOVAs under the four designed interaction scenarios.
A total of 32 participants (15 males and 17 females) with an age range of 23 to 68 years (mean: 49.12, standard deviation: 11.13) participated in this experiment as pedestrians.
Informed consent was obtained from each participant.

\subsection{Experimental Car and eHMI Design}

\begin{figure}[htb]
\centering
\includegraphics[width=0.8\linewidth]{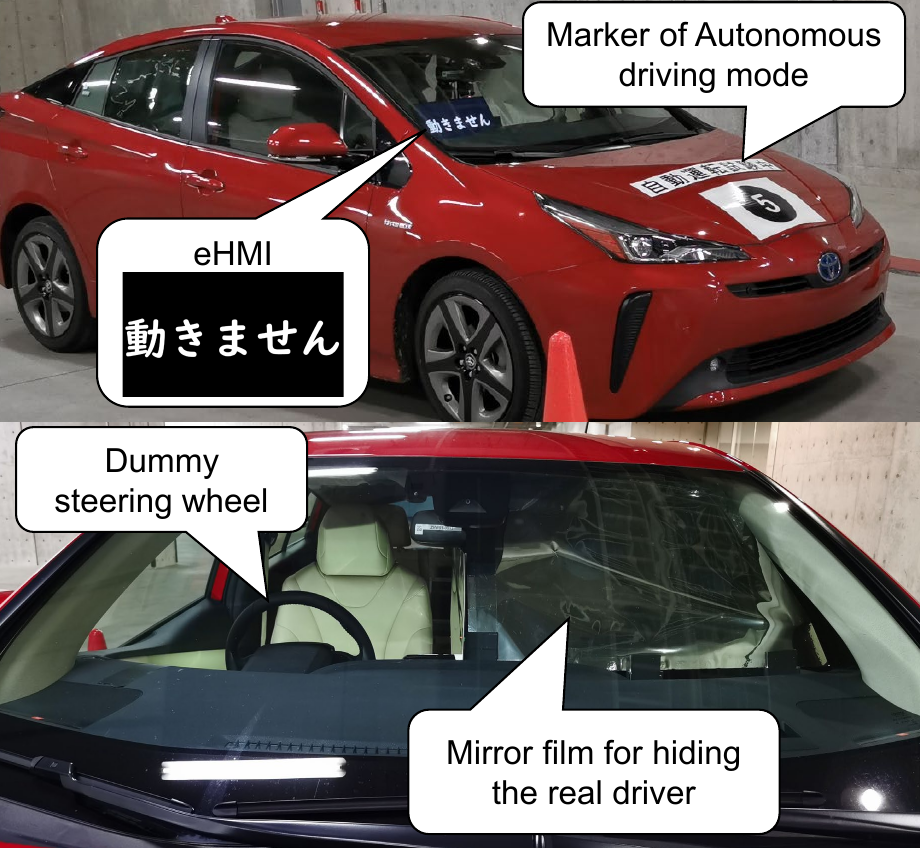}
\caption{Experimental car with an eHMI: a left-hand-driven car stimulating a right-hand-driven AV. An eHMI is installed on the right bottom of the windshield.}
\label{fig:exp_car}
\end{figure}

To achieve the ghost driver trick, a left-hand drive car (Toyota Prius) was used to simulate an AV (see Fig.~\ref{fig:exp_car}) because cars in Japan are typically right-hand drive.
In the left seat of the experimental car, the expert driver was hidden by a mirror film. 
Additionally, a dummy steering wheel was installed on the right side to ensure that the participants assume that the experimental car was a right-hand-drive car.
The maximum speed of the car was limited to 8~km/h and the average speed was approximately 4~km/h.
Particularly, this speed considered the experimental scene of the parking lot and the safety of the participants.

An eHMI display was installed on the right rear of the windshield (see the upper part of Fig.~\ref{fig:exp_car}).
The message ``\begin{CJK}{UTF8}{ipxm}動きません\end{CJK}'' (``\textit{UGOKIMASEN}'': the AV is stationary at that this moment) appears immediately on the eHMI after the AV stops. 
After a pedestrian crosses the road, the eHMI blinks twice within two seconds (1~Hz), indicating that the car will depart.
Following the blinks, the eHMI turns off, and the car departs.

\begin{figure}[tb]
\centering
\includegraphics[width=1\linewidth]{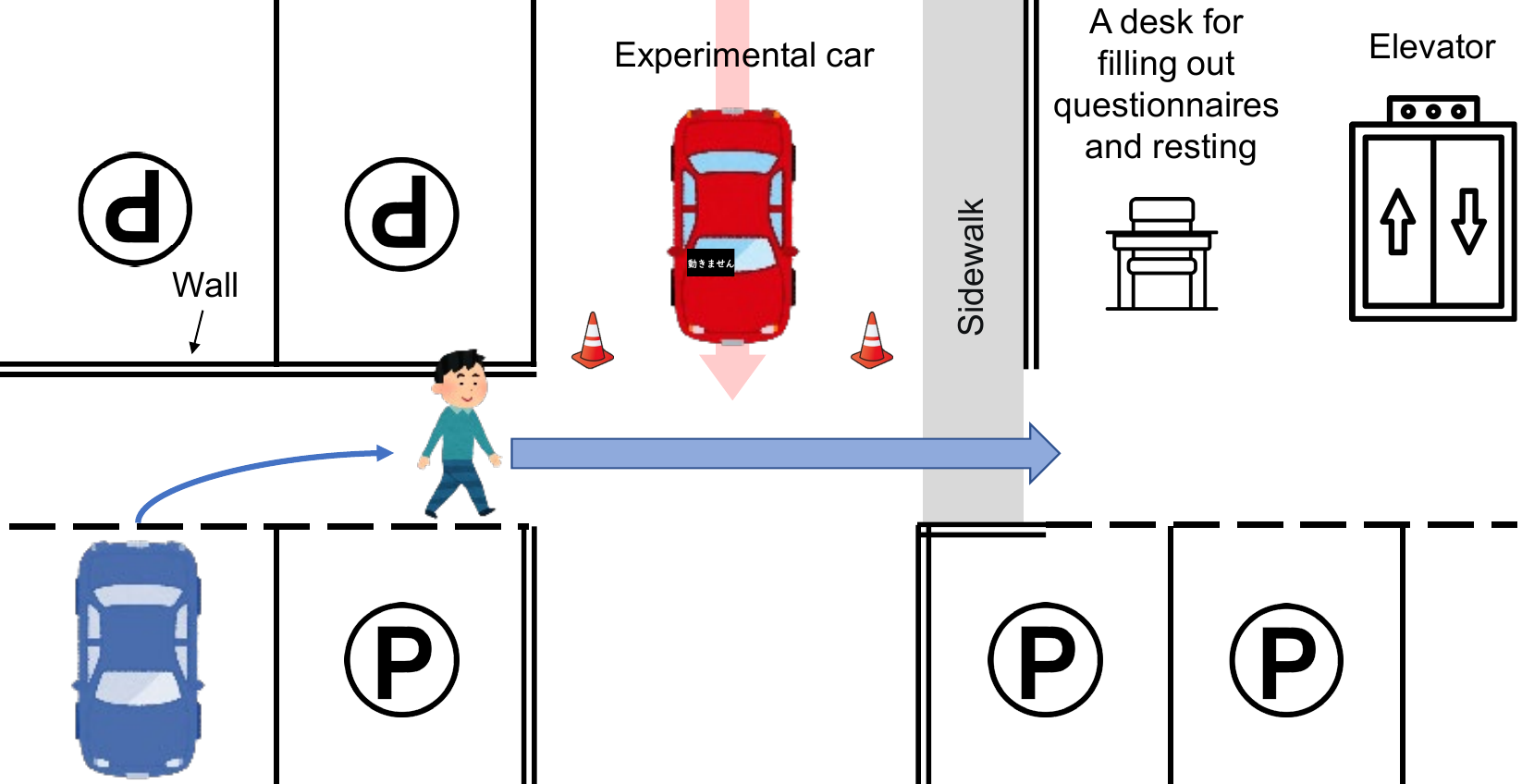}
\caption{Experimental scene: simulation of pedestrian encounters with a car in a parking lot.}
\label{fig:map}
\end{figure}

There are two reasons to consider the above configuration:
(1) To eliminate potential accident liabilities when eHMIs are practically implemented in the future, eHMIs will not issue directives to pedestrians.
``\textit{UGOKIMASEN}'' only informs pedestrians about the AV's current driving state and intentions; it does not instruct them on what to do, such as ``You go first.'' 
The pedestrians need to decide their walking behavior based on the state of the car and the message on the eHMI. 
They are responsible for their decisions and behaviors.
(2) Although text-based eHMI convey clear information that may be perceived as not requiring the pre-instruction, this message ``\textit{UGOKIMASEN}'' provides pedestrians with a vague understanding of the AV's intention in the specific Japanese context. 
Thus, this message can help to compare the effectiveness of the pre-instruction in this experiment.
For example, a pedestrian might think that the AV is asking for help because the car is out of order when the ``\textit{UGOKIMASEN}'' message appears after the car has stopped.
In addition, the blinking conditions and timing of the message remain unclear to the pedestrians before the pre-instruction. 
Thus, this ambiguity presents an important reason for instructing pedestrians.

\subsection{Wizard of Oz Experiment Introductions for Participants}

The following information was introduced to the participants before the experiment commenced:
\begin{enumerate}
\item Please imagine that you are driving to the shopping mall and parking your car in the underground parking garage.
\item After parking, you will cross the road to reach the elevator.
Please walk at a normal speed during this process.
\item When you cross the road, a manual driving car (\ie MV) or an automated driving car (\ie AV) will arrive. 
You should be mindful of it when crossing the road.
\item The AV is a driverless car with multiple built-in advanced sensors that can detect the surrounding environment, such as pedestrians, roads, and stop lines. \textbf{(fake information)}
\item The pylon marks the stop line (see Fig.~\ref{fig:map}). The MV or AV will stop before the stop line. 
Subsequently, the vehicle will decide whether to depart from the surrounding situation based on the presence of a pedestrian.
\end{enumerate}

Notably, any information about eHMI was withheld from the participants in this study before receiving the introduction. 
Furthermore, the AV interaction scenarios described in the following subsection were not explained to the participants so they could not easily predict the content of each trial.

\subsection{Interaction Scenarios}

\begin{figure*}[h]
\centering
\includegraphics[width=0.9\linewidth]{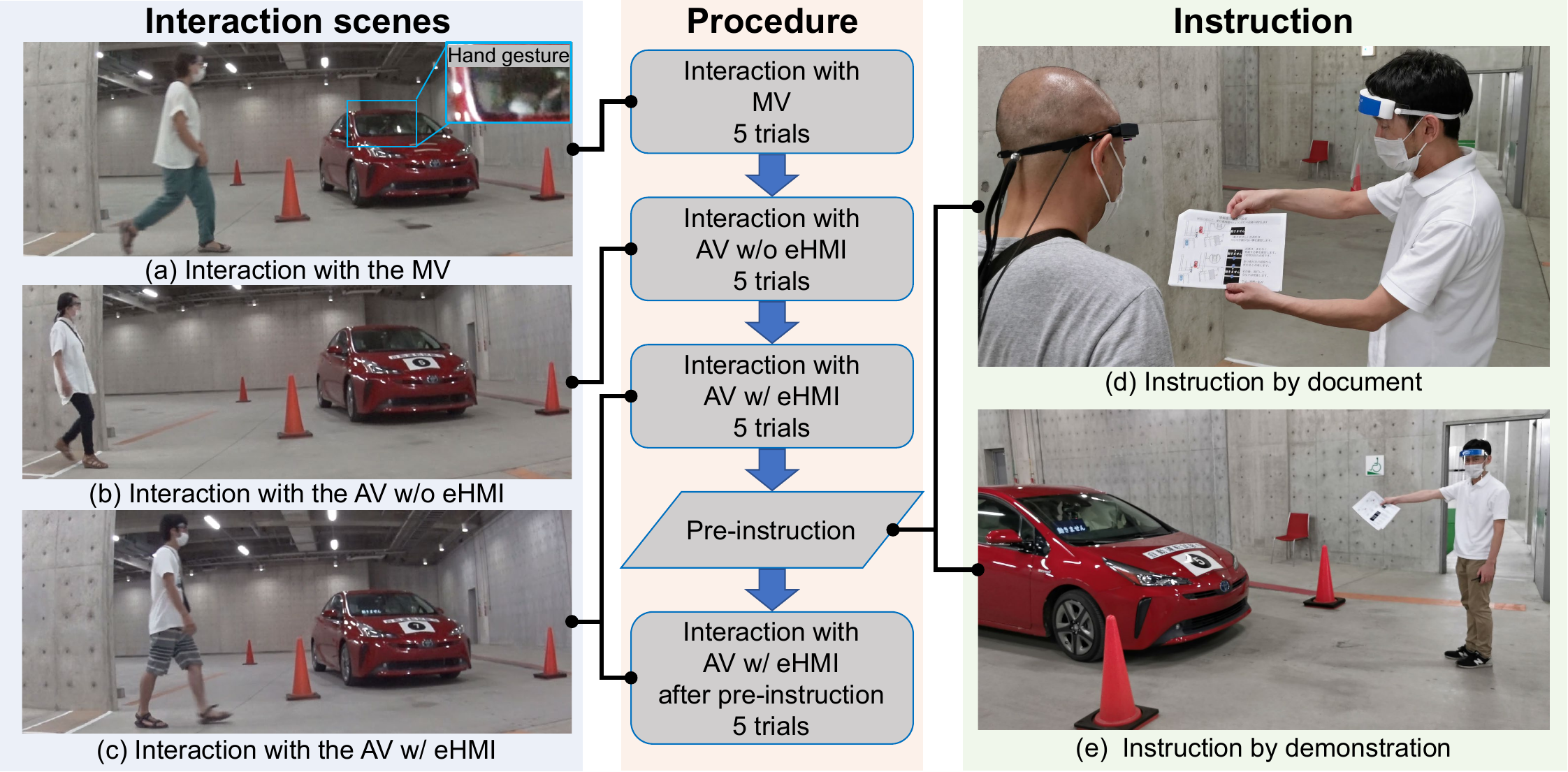}
\caption{Procedure of a within-participants design experiment.}
\label{fig:procedure}
\end{figure*}

\begin{figure*}[h!]
\centering
\includegraphics[width=0.8\linewidth]{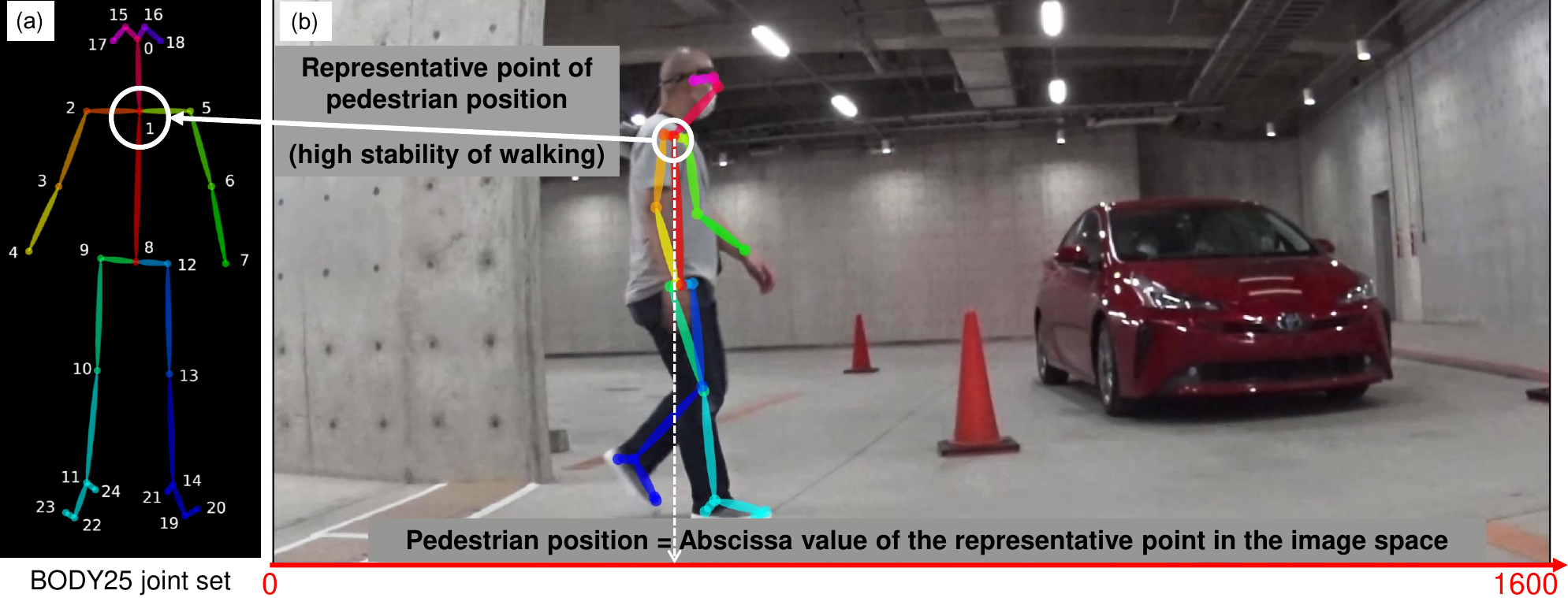}
\caption{The walking behaviors of participants are estimated using OpenPose with BODY~25 joint set. (a) shows the BODY~25 joint set. (b) Method for extracting the representative point of pedestrian position.}
\label{fig:openpose}
\end{figure*}

To compare pedestrians' subjective feelings and walking behavior changes with and without the use of eHMI, as well as before and after receiving pre-instructions, a road crossing experiment using a within-subject design was conducted.
Fig.~\ref{fig:procedure} reveals that the pedestrians interact with the experimental car in four designed scenarios: \textbf{\textit{MV}}, \textbf{\textit{AV w/o eHMI}}, \textbf{\textit{AV w/ eHMI}}, and \textbf{\textit{AV w/ eHMI after PI}} (\ie \textbf{\textit{AV w/ eHMI after pre-instuction}}).

\textbf{\textit{ MV:}}
The first scenario examines an interaction where a pedestrian encounters an MV (see Fig.~\ref{fig:procedure}~(a)). 
The dummy driver sits in the right seat and holds the dummy steering wheel.
While the real driver controls the experimental car in the left seat covered by the mirror film. When a pedestrian encounters the MV, the pedestrian can see a dummy driver driving the experimental car. 
Furthermore, the dummy driver yields the right-of-way to the pedestrian after stopping the car by using a typical Japanese hand gesture of ``After You'', that is, \ie moving the driver's hand forward once with palm facing upwards (looks like \faIcon{hand-holding}).
The MV does not depart until the pedestrian has crossed the road.

\textbf{\textit{AV w/o eHMI}:}
The second scenario examines an interaction where a pedestrian encounters an AV without the eHMI (see Fig.~\ref{fig:procedure}~(b)).
Notably, no driver sits in the right seat.
The real driver controls the experimental car in the left seat covered by the mirror film. 
No eHMI device is present in the AV but a sign on the hood indicates the autonomous mode (see Fig.~\ref{fig:exp_car}). 
The AV stops before the stop line (two pylons) when it encounters the pedestrian. 
At this time, the pedestrian has to decide how and when to cross the road. 
Finally, the AV starts moving after the pedestrian has completely crossed the road.

\textbf{\textit{AV w/ eHMI}:} 
This third scenario examines an interaction where a pedestrian encounters an AV that has the eHMI (see Fig.~\ref{fig:procedure}~(c)).
The configuration of the driver is consistent with \textit{AV w/o eHMI}.
Particularly, the difference is that an eHMI device is installed behind the right side of the windshield. The AV stops before the stop line and the eHMI displays a message ``\textit{UGOKIMASEN}'' informing the pedestrian that the AV is stationary. Subsequently, the pedestrian needs to decide how and when to cross the road.
The eHMI message blinks twice when the pedestrian has completely crossed the road. Lastly, the eHMI is turned off and the AV departs.

\textbf{\textit{AV w/ eHMI after PI}:} 
The fourth scenario is consistent with \textit{AV w/ eHMI} (see Fig.~\ref{fig:procedure}~(c)). 
The difference is that two types of pre-instructions are used to help the pedestrian adjust the mental models of the AV with the eHMI before the trials of this scenario.
Specifically, to eliminate the participants' vague understanding of the driving intentions of the AV (\textit{comprehension}) and to help them to predict the driving behavior (\textit{projection}) in situation awareness, a document (see Fig.~\ref{fig:procedure}~(d)) explaining the meaning of the information on the eHMI is presented based on the following information:
\begin{enumerate}
\item When the AV detects the pedestrian, the message ``\textit{UGOKIMASEN}'' appears on the eHMI after the car stops (for \textit{projection}). 
\item ``\textit{UGOKIMASEN}'' indicating that the AV does not move (for \textit{comprehension}).
\item Thereafter, the eHMI will blink twice in two seconds after the pedestrian crosses the road (for \textit{projection}). 
\item The blinking indicates that the AV will move again and depart (for \textit{comprehension}).
\item After blinking, the eHMI is turned off. Subsequently, the AV will move (for \textit{projection}).
\end{enumerate}
Meanwhile, after the document-based pre-instruction, a demonstration (see Fig.~\ref{fig:procedure}~(e)) is also used to explain to the pedestrians when the eHMI would be turned on/off and eHMI would blink.
To conceal the ghost driver trick, pedestrians stood by the side of the road to watch the demonstrations. After the two types of pre-instruction described above, pedestrians are allowed to ask additional questions if they have doubts about the content of the pre-instruction. The fourth scenario, that is {\textit{AV w/ eHMI after PI}}, is performed after confirming that the pedestrians understood the contents of the pre-instruction.

Note that in all four scenarios above, if the pedestrian hesitates before crossing the road or yields to the vehicle, \ie MV or AV, the vehicle will maintain its stopping until the pedestrian completes their crossing of the road.

\subsection{Experimental Procedure}
\label{sec:order}

In the experiment, each scenario was run five times. 
In total, each participant encountered the experimental car across 20 trials. After crossing the road in each trial, the participants were asked to sit on a chair (see Fig.~\ref{fig:map}) to complete the questionnaires and rest for approximately one minute. In particular, the purpose of this rest is to prevent the effect of fatigue on walking behavior.

The sequence of the scenarios and the pre-instruction for each participant is displayed in the central part of Fig.~\ref{fig:procedure}, that is, \textit{MV} $\rightarrow$ \textit{AV w/o eHMI} $\rightarrow$ \textit{AV w/ eHMI} $\rightarrow$ \textit{pre-instruction} $\rightarrow$\textit{AV w/ eHMI after PI}.
The experiment did not use a random order due to the following reasons.
Starting with \textit{MV} first aimed to familiarize participants with encountering vehicles at the test site and establish a baseline for subjective evaluations.
Next, \textit{AV w/ eHMI} followed \textit{AV w/o eHMI} to prevent participants from mistakenly thinking that the eHMI is corrupted in \textit{AV w/o eHMI} scenario.
Furthermore, \textit{AV w/ eHMI after PI} has to appear after \textit{AV w/ eHMI} to conform to the current experiment’s purpose, that is, comparing the differences in participants' subjective evaluations of the eHMI-equipped AV and walking behavior before and after the pre-instruction.
More significantly, the above order complies with the order of the popularization of MV and AV, as well as eHMI.

\subsection{Measurements}
This study measured participants' subjective feelings and walking behaviors during encounters with the experimental car in the four scenarios.

\subsubsection{Subjective evaluations}
\label{sec:SE_Q}
Based on the cognition-decision-behavior model~\citep{Liu2022_APMV} (see Fig.~\ref{fig:model}), six questions in Japanese were designed to evaluate the participants' subjective feelings as follows:
\begin{enumerate}
\small
\item[Q1:] Was it easy to understand the driving intention of the car?
\item[Q2:] Was it easy to predict the behavior of the car?
\item[Q3:] Did you feel the behavior of the car was dangerous?
\item[Q4:] Did you trust the car when you crossed the road?
\item[Q5:] Did you feel a sense of relief when you crossed the road?
\item[Q6:] Did you hesitate when you crossed the road?
\end{enumerate}
These questions were validated for their effectiveness through causal discovery in \citep{liu_IV24}.
As shown in Fig.~\ref{fig:model}, Q1 and Q2 are used to evaluate the comprehension and projection steps in the situation model. 
The Q3, Q4, and Q5 are used for risk evaluation while Q6 is used to evaluate the difficulty with making a decision to cross the road.
After each trial, the participants were asked to answer the above six questions on a 5-point scale: ``\textit{1=Strongly Disagree}'', ``\textit{2=Disagree}'', ``\textit{3=Undecided}'', ``\textit{4= Agree}'' and ``\textit{5=Strongly Agree}''.

\subsubsection{Walking behavior}

The walking behaviors of the participants were recorded on video (1600x900 pixels with 60~FPS) from the side using a camera installed on the road (see Fig.~\ref{fig:openpose}).
The positions of the participants during the crossings were calculated using OpenPose~\citep{openPose}. 
OpenPose detects skeletal feature points based on the BODY~25 joint set (Fig.~\ref{fig:openpose}~(a)) in each frame. 
The position of pedestrians is subsequently determined using the neck feature point along the horizontal axis of the image space, denoted as $x_t$ (where $t$ is the frame number), as the neck movement is minimally affected by walking motion (Fig.~\ref{fig:openpose}~(b)).
For accuracy, we used detected positions $x_t$ within the range of [50 pixels, 1550 pixels].
The participants' walking speeds were calculated based on the positional difference between consecutive frames, \ie $s_t= (x_t-x_{t-1})\times 60~[\rm{pixels/second}]$, with 60 FPS videos.

\section{RESULTS}

\subsection{Subjective evaluations}

\begin{figure}[bh]
\centering
\includegraphics[width=1\linewidth]{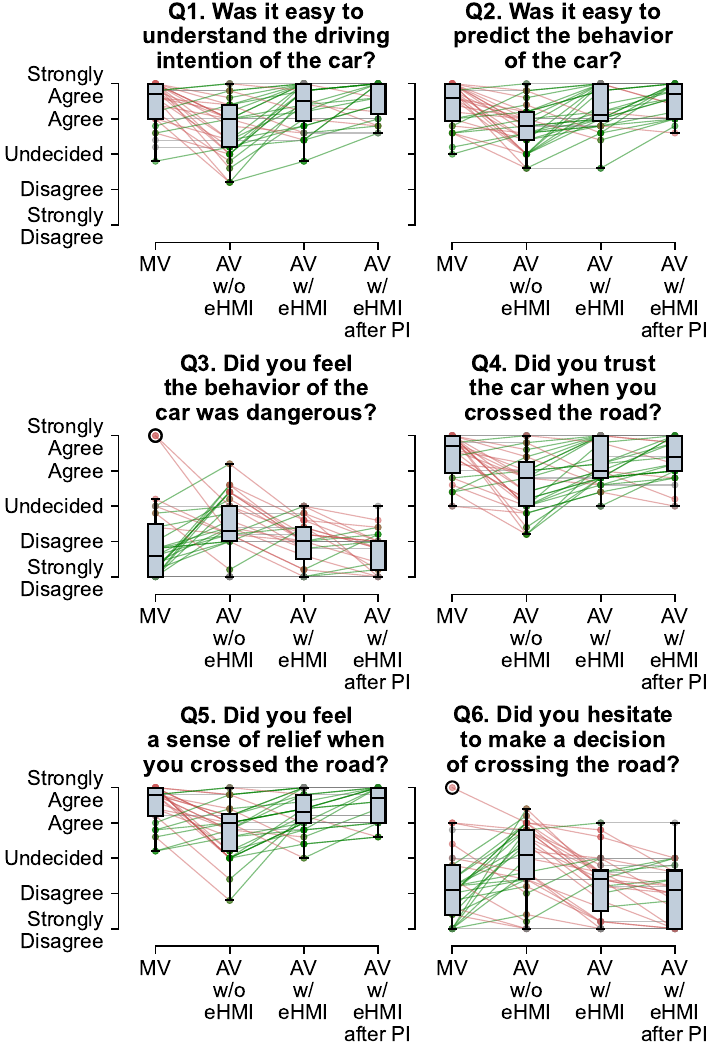}
\caption{Boxplots and paired poinplots were used to visualize the subjective evaluation of six questions across four scenarios by 32 participants.
Paired pointplots showed evaluated scores changes under the four scenarios for each participant, with indicated greater agreement with the question, while red lines indicated greater disagreement.
}
\label{fig:result_01}
\end{figure}

In Fig.~\ref{fig:result_01}, boxplots and paired pointplos were employed to visualize the subjective evaluation results of Q1-Q6 for four scenarios, as reported by 32 participants.
Paired pointplots showed evaluated scores changes under the four scenarios for each participant, with indicated greater agreement with the question, while red lines indicated greater disagreement.

\begin{table}[h!b]
\renewcommand{\arraystretch}{1}
\centering
\caption{Friedman F-test for Q1 to Q6 among four interaction scenarios. ***: $p<0.001$.}
\label{tab:QA_Friedman}
\renewcommand{\arraystretch}{1}
\setlength{\tabcolsep}{4mm}
\begin{tabular}{@{}llllrl@{}}
\toprule
 & \multicolumn{1}{c}{\textit{W}} & \multicolumn{1}{c}{ddof1} & \multicolumn{1}{c}{ddof2} & \multicolumn{1}{c}{\textit{F}} & \multicolumn{1}{c}{\textit{p}} \\ \midrule
Q1 & 0.225 & 2.938 & 91.062 & 8.976 & $<0.001$ *** \\
Q2 & 0.255 & 2.938 & 91.062 & 10.612 & $<0.001$ *** \\
Q3 & 0.228 & 2.938 & 91.062 & 9.173 & $<0.001$ *** \\
Q4 & 0.349 & 2.938 & 91.062 & 16.647 & $<0.001$ *** \\
Q5 & 0.273 & 2.938 & 91.062 & 11.657 & $<0.001$ *** \\
Q6 & 0.193 & 2.938 & 91.062 & 7.415 & $<0.001$ *** \\ \bottomrule
\end{tabular}
\end{table}

\begin{table}[h!b]
\centering
\footnotesize
\caption{Post hoc multiple comparisons via two-sided Wilcoxon signed-rank test~(WSR) with the Benjamini/Hochberg FDR correction for Q1 to Q6 among four interaction scenario. \\ *: $p<0.05$, **: $p<0.01$, ***: $p<0.001$.}
\label{tab:QA_Posthoc}
\renewcommand{\arraystretch}{1}
\setlength{\tabcolsep}{1.4mm}{
\begin{tabular}{@{}cllrlr@{}}
\toprule
& Scenario A  & Scenario B & \multicolumn{1}{c}{\textit{W}} & \multicolumn{1}{c}{\textit{p-adj.}} & \textit{hedges g} \\ \midrule
\multirow{6}{*}{Q1} & MV & AV w/o eHMI & 30.000 & 0.006 ** & 0.786 \\
 & MV & AV w/ eHMI & 32.500 & 0.336 & 0.163 \\
 & MV & AV w/ eHMI after PI & 30.000 & 0.312 & -0.242 \\
 & AV w/o eHMI & AV w/ eHMI & 30.500 & 0.030 * & -0.640 \\
 & AV w/o eHMI & AV w/ eHMI after PI & 5.500 & 0.001 *** & -1.127 \\
 & AV w/ eHMI & AV w/ eHMI after PI & 4.000 & 0.037 * & -0.440 \\ \midrule
\multirow{6}{*}{Q2} & MV & AV w/o eHMI & 30.000 & 0.005 ** & 0.908 \\
 & MV & AV w/ eHMI & 34.000 & 0.059 & 0.393 \\
 & MV & AV w/ eHMI after PI & 42.000 & 0.488 & -0.165 \\
 & AV w/o eHMI & AV w/ eHMI & 31.000 & 0.034 * & -0.540 \\
 & AV w/o eHMI & AV w/ eHMI after PI & 8.000 & 0.001 *** & -1.136 \\
 & AV w/ eHMI & AV w/ eHMI after PI & 7.000 & 0.005 ** & -0.598 \\ \midrule
\multirow{6}{*}{Q3} & MV & AV w/o eHMI & 59.500 & 0.039 * & -0.528 \\
 & MV & AV w/ eHMI & 57.000 & 0.632 & -0.097 \\
 & MV & AV w/ eHMI after PI & 38.500 & 0.632 & 0.134 \\
 & AV w/o eHMI & AV w/ eHMI & 20.000 & 0.021 * & 0.558 \\
 & AV w/o eHMI & AV w/ eHMI after PI & 7.500 & 0.001 *** & 0.863 \\
 & AV w/ eHMI & AV w/ eHMI after PI & 0.000 & 0.021 * & 0.329 \\ \midrule
\multirow{6}{*}{Q4} & MV & AV w/o eHMI & 21.500 & 0.003 ** & 1.022 \\
 & MV & AV w/ eHMI & 5.000 & 0.022 * & 0.417 \\
 & MV & AV w/ eHMI after PI & 16.500 & 0.227 & 0.179 \\
 & AV w/o eHMI & AV w/ eHMI & 24.500 & 0.006 ** & -0.686 \\
 & AV w/o eHMI & AV w/ eHMI after PI & 18.500 & 0.002 ** & -0.855 \\
 & AV w/ eHMI & AV w/ eHMI after PI & 10.000 & 0.132 & -0.226 \\ \midrule
\multirow{6}{*}{Q5} & MV & AV w/o eHMI & 28.000 & 0.003 ** & 0.970 \\
 & MV & AV w/ eHMI & 34.000 & 0.033 * & 0.494 \\
 & MV & AV w/ eHMI after PI & 49.000 & 0.835 & 0.052 \\
 & AV w/o eHMI & AV w/ eHMI & 30.500 & 0.032 * & -0.542 \\
 & AV w/o eHMI & AV w/ eHMI after PI & 20.500 & 0.001 *** & -0.955 \\
 & AV w/ eHMI & AV w/ eHMI after PI & 6.000 & 0.015 * & -0.463 \\ \midrule
\multirow{6}{*}{Q6} & MV & AV w/o eHMI & 36.000 & 0.010 ** & -0.648 \\
 & MV & AV w/ eHMI & 56.500 & 0.854 & -0.030 \\
 & MV & AV w/ eHMI after PI & 45.500 & 0.477 & 0.122 \\
 & AV w/o eHMI & AV w/ eHMI & 26.000 & 0.008 ** & 0.651 \\
 & AV w/o eHMI & AV w/ eHMI after PI & 15.000 & 0.001 *** & 0.816 \\
 & AV w/ eHMI & AV w/ eHMI after PI & 8.000 & 0.228 & 0.162 \\   \bottomrule
\end{tabular}
}
\end{table}

Afterwards, the Shapiro-Wilk test indicated that the results of Q1-Q6 under the four eHMI conditions departed significantly from normality, respectively.
Therefore, Friedman F-test was performed and indicated that there were significant differences among the four scenarios for each subjective evaluation question, which were shown in Table~\ref{tab:QA_Friedman}.
Therefore, post-hoc multiple comparisons were performed between each pair of the four scenarios for each question using a two-sided Wilcoxon signed-rank test~(WSR) with the Benjamini/Hochberg FDR correction.
The results for Q1-Q6 are shown in Table~\ref{tab:QA_Posthoc}.
It shows that the participants' subjective feelings had significant differences across the various scenarios. Details of these findings are discussed in Section~\ref{sec:Dis_Q}.

\subsection{Walking behaviors before and after stopping the car}

\begin{figure}[h!p]
\centering
\includegraphics[width=1\linewidth]{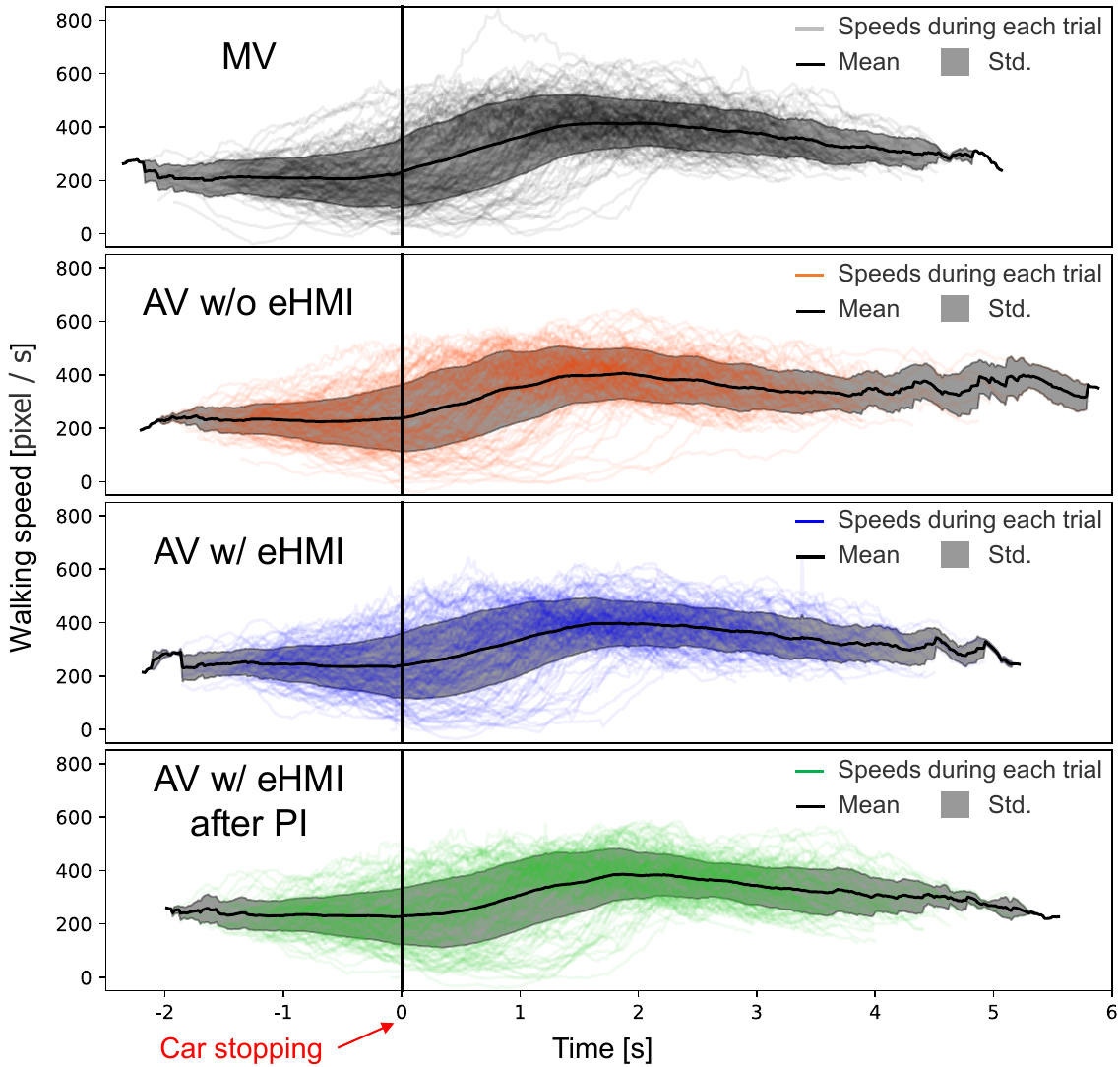}
\caption{Walking speeds of 27 pedestrians during 5 trials of four scenarios. The car stopping time is the 0 second.}
\label{fig:walk_speed}

\vspace{2mm}

\footnotesize
\captionof{table}{The means and standard deviations (Std.) of walking speeds in all trials per 0.2 seconds interval. The car stopping time is the 0 second. The maximum and minimum values are shown by bold and underline, separately.}
\label{Tab:mean_std_speed}
\renewcommand{\arraystretch}{1}
\setlength{\tabcolsep}{1mm}{
\begin{tabular}{@{}c|cccc|cccc@{}}
\toprule
 & \multicolumn{4}{c|}{Mean of walking speeds [pixels/s]} & \multicolumn{4}{c}{Std. of walking speeds [pixels/s]} \\ \cmidrule(l){2-9} 
\begin{tabular}[c]{@{}c@{}}Time\\ {[}s{]}\end{tabular} & \multicolumn{1}{c}{MV} & \multicolumn{1}{c}{\begin{tabular}[c]{@{}c@{}}AV\\ w/o\\ eHMI\end{tabular}} & \multicolumn{1}{c}{\begin{tabular}[c]{@{}c@{}}AV\\ w/\\ eHMI\end{tabular}} & \multicolumn{1}{c|}{\begin{tabular}[c]{@{}c@{}}AV w/\\ eHMI\\ after  PI\end{tabular}} & \multicolumn{1}{c}{MV} & \multicolumn{1}{c}{\begin{tabular}[c]{@{}c@{}}AV\\ w/o\\ eHMI\end{tabular}} & \multicolumn{1}{c}{\begin{tabular}[c]{@{}c@{}}AV\\ w/\\ eHMI\end{tabular}} & \multicolumn{1}{c}{\begin{tabular}[c]{@{}c@{}}AV w/\\ eHMI\\ after  PI\end{tabular}} \\ \midrule

-1.0 & {\ul 210.3} & 231.5 & \textbf{245.5} & 230.6 & \textbf{72.82 }& 60.86 & 61.94 & {\ul 56.13} \\
-0.8 & {\ul 208.2} & 226.1 & \textbf{244.7} & 234.9 & \textbf{79.33 }& 73.30 & 69.96 & {\ul 63.47} \\
-0.6 & {\ul 206.9} & 225.6 & \textbf{236.0} & 231.3 & \textbf{91.18} & 83.76 & 77.87 & {\ul 71.25} \\
-0.4 & {\ul 207.4} & 228.7 & \textbf{235.0} & 231.1 & \textbf{103.8 }& 95.55 & 88.13 & {\ul 82.20} \\
-0.2 & {\ul 216.4} & 232.8 & \textbf{236.8} & 229.9 &\textbf{ 112.9 }& 110.5 & 103.6 & {\ul 90.35} \\
0.0 & 229.7 & 239.1 & \textbf{239.2} & {\ul 228.3} & \textbf{128.6 }& 123.8 & 120.8 & {\ul 105.8} \\
+0.2 & 258.4 & \textbf{258.8} & 254.3 & {\ul 234.9} & \textbf{141.5}& 137.9 & 137.5 & {\ul 118.5} \\
+0.4 & 280.7 & \textbf{283.1} & 268.2 & {\ul 242.5} & \textbf{144.0} & 142.4 & 141.7 & {\ul 127.4} \\
+0.6 & \textbf{311.5} & 303.5 & 289.9 & {\ul 256.5} & \textbf{151.1} & 139.9 & 144.1 & {\ul 129.1} \\
+0.8 & 336.9 & \textbf{339.1} & 311.4 & {\ul 280.5} & \textbf{151.2} & 144.0 & 140.3 & {\ul 132.5} \\
+1.0 & \textbf{362.4} & 355.0 & 335.9 & {\ul 307.8} & \textbf{140.2} & 131.1 & 133.4 & {\ul 129.0} \\
+1.2 & \textbf{388.1} & 373.4 & 361.0 & {\ul 333.5} & \textbf{130.2} & 122.9 & {\ul 119.5} & 125.4 \\
+1.4 & \textbf{407.3} & 396.4 & 381.7 & {\ul 349.5} &  113.3 & 108.4 & {\ul 106.1} & \textbf{117.5} \\
+1.6 & \textbf{413.6} & 397.0 & 396.1 & {\ul 364.6} &  101.0 & 98.58 &{\ul96.11} & \textbf{108.4} \\
+1.8 & \textbf{415.1} & 405.8 & 396.8 & {\ul 384.5} & 91.68 & 93.95 & {\ul84.71} & \textbf{96.66} \\
+2.0 & \textbf{415.0} & 402.4 & 396.6 & {\ul 382.2} & 84.94 & \textbf{94.50} & {\ul 82.66} & 84.87 \\
+2.2 & \textbf{414.5} & 388.9 & 398.1 & {\ul 383.2} & 81.89 & \textbf{89.53} & 78.38 & {\ul 77.41} \\
+2.4 & \textbf{406.4} & 386.5 & 393.7 & {\ul 376.4} & 85.12 & \textbf{87.83} & 72.13 & {\ul 71.66} \\
+2.6 & \textbf{401.2} & 379.7 & 386.2 & {\ul 368.5} & 78.65 & \textbf{79.31} & 70.83 & {\ul 61.94} \\
+2.8 & \textbf{392.8} & 371.5 & 379.4 & {\ul 360.4} & 77.03 & \textbf{79.80} & 70.49 & {\ul 63.91} \\
+3.0 & \textbf{390.4} & 359.9 & 374.3 & {\ul 350.4} & 72.53 & \textbf{78.03} & 71.80 & {\ul 70.28} \\ \bottomrule
\end{tabular}
}

\vspace{2mm}

\centering
\footnotesize
\captionof{table}{Effects of time, scenarios and their interaction on walking speeds via a repeated measured two-way ANOVA (two-sided). *:$p<0.05$, **:$p<0.01$, ***:$p<0.001$.}

\label{Tab:2way_ANOVA}
\setlength{\tabcolsep}{1.2mm}
\begin{tabular}{c|rrrrlrr}
\toprule
Source & \multicolumn{1}{c}{ddof1} & \multicolumn{1}{c}{ddof2} & \multicolumn{1}{c}{\textit{F}} & \multicolumn{2}{c}{\textit{p-GG-corr}} & \multicolumn{1}{c}{$\eta_g^2$} & \multicolumn{1}{c}{$\epsilon$-GG} \\ \midrule
Time & 20 & 300 & 26.787 & $<0.001$ & *** & 0.514 & 0.102 \\
Scenarios & 3 & 45 & 5.250 & 0.010 & ** & 0.008 & 0.700 \\ 
Time$\times$Scenarios & 60 & 900 & 4.245 & 0.024 & * & 0.028 & 0.033 \\ \bottomrule
\end{tabular}
\end{figure}

Unfortunately, due to technical issues, only the videos of 27 out of the 32 participants crossing the road were successfully saved.
To analyze the walking behaviors of the 27 participants before and after the car stopped, this study regarded the moment the car stopping as 0 on the timeline in each trial. 
Thus, the walking speeds of these 27 participants were calculated from the videos using OpenPose as shown in Fig.~\ref{fig:walk_speed}.
In which, we found a difference in the timings for the participants to enter and exit the crossing scene due to the difference in their walking speed.
Therefore, walking speeds from 1 second before to 3 seconds after the car stopped, that is [-1.0, 3.0] seconds, were selected for analysis to address this issue.
In addition, to analyze the differences in the walking speed of the participants at each moment in each scenario, the representative time points for walking speed were extracted in 5~[Hz].
Meanwhile, to examine that the pre-instruction of eHMI can better perform the guidance effect of eHMI on pedestrian walking behavior, the standard deviation~(Std.) of the participants' walking speeds were assessed at each time point.

\begin{table*}[htb]
\centering
\caption{Post-hoc pairwise comparisons of four scenarios effected on walking speeds by paired t-tests (two-sided) with Bonferroni correction. *:$p<0.05$.}
\label{Tab:post-hoc_eHMI}
\begin{tabular}{@{}llccccrclr@{}}

\toprule
\multicolumn{1}{c}{Scenarios A} &
  \multicolumn{1}{c}{scenarios B} &
  mean(A) &
  std(A) &
  mean(B) &
  std(B) &
  \multicolumn{1}{c}{\textit{T}} &
  dof &
  \multicolumn{1}{c}{\textit{p-adj}} &
  \textit{cohen}\\ \midrule
MV          & AV w/o eHMI         & 300.6 & 41.4 & 294.8 & 49.7 & 0.922  & 15 & 1.000   & 0.128  \\
MV          & AV w/ eHMI          & 300.6 & 41.4 & 295.7 & 47.9 & 1.054  & 15 & 1.000   & 0.111  \\
MV          & AV w/ eHMI after PI & 300.6 & 41.4 & 280.7 & 51.0 & 2.902  & 15 & 0.066   & 0.430   \\
AV w/o eHMI & AV w/ eHMI          & 294.8 & 49.7 & 295.7 & 47.9 & -0.292 & 15 & 1.000   & -0.018 \\
AV w/o eHMI & AV w/ eHMI after PI & 294.8 & 49.7 & 280.7 & 51.0 & 2.582  & 15 & 0.125   & 0.281  \\
AV w/ eHMI  & AV w/ eHMI after PI & 295.7 & 47.9 & 280.7 & 51.0 & 3.379  & 15 & 0.025 * & 0.303  \\ \bottomrule

\end{tabular}
\end{table*}

\begin{table*}[h]
\centering
\footnotesize
\caption{Simple main effects of four scenarios within time points effected on walking speeds by paired t-tests (two-sided) with Bonferroni correction. *:$p<0.05$, **:$p<0.01$, ***:$p<0.001$.}
\label{Tab:post-hoc}
\setlength{\tabcolsep}{0.6mm}
\begin{tabular}{@{}c|clcclcclcclcclcclc@{}}
\toprule
\multirow{2}{*}{\begin{tabular}[c]{@{}c@{}}Time\\ {[}s{]}\end{tabular}} & \multicolumn{3}{c|}{\begin{tabular}[c]{@{}c@{}}MV\\ v.s.\\ AV w/o eHMI\end{tabular}} & \multicolumn{3}{c|}{\begin{tabular}[c]{@{}c@{}}MV\\ v.s.\\ AV w/ eHMI\end{tabular}} & \multicolumn{3}{c|}{\begin{tabular}[c]{@{}c@{}}MV\\ v.s.\\ AV w/ eHMI\\ after PI\end{tabular}} & \multicolumn{3}{c|}{\begin{tabular}[c]{@{}c@{}}AV w/o eHMI\\ v.s.\\ AV w/ eHMI\end{tabular}} & \multicolumn{3}{c|}{\begin{tabular}[c]{@{}c@{}}AV w/o eHMI\\ v.s.\\ AV w/ eHMI\\ after PI\end{tabular}} & \multicolumn{3}{c}{\begin{tabular}[c]{@{}c@{}}AV w/ eHMI\\ v.s.\\ AV w/ eHMI\\ after PI\end{tabular}} \\ \cmidrule(l){2-19} 
 & \textit{T} & \multicolumn{1}{c}{\textit{p-adj}} & \multicolumn{1}{c|}{\textit{cohen}} & \textit{T} & \multicolumn{1}{c}{\textit{p-adj}} & \multicolumn{1}{c|}{\textit{cohen}} & \textit{T} & \multicolumn{1}{c}{\textit{p-adj}} & \multicolumn{1}{c|}{\textit{cohen}} & \textit{T} & \multicolumn{1}{c}{\textit{p-adj}} & \multicolumn{1}{c|}{\textit{cohen}} & \textit{T} & \multicolumn{1}{c}{\textit{p-adj}} & \multicolumn{1}{c|}{\textit{cohen}} & \textit{T} & \multicolumn{1}{c}{\textit{p-adj}} & \textit{cohen} \\ \midrule
-1.0 & -2.564 & 0.104 & \multicolumn{1}{c|}{-0.376} & -4.749 & 0.001 *** & \multicolumn{1}{c|}{-0.628} & -2.734 & 0.071 & \multicolumn{1}{c|}{-0.361} & -3.742 & 0.006 ** & \multicolumn{1}{c|}{-0.248} & 0.346 & 1.000 & \multicolumn{1}{c|}{0.031} & 3.032 & 0.036 * & 0.289 \\
-0.8 & -1.744 & 0.558 & \multicolumn{1}{c|}{-0.216} & -4.047 & 0.002 ** & \multicolumn{1}{c|}{-0.503} & -2.783 & 0.059 & \multicolumn{1}{c|}{-0.348} & -3.567 & 0.009 ** & \multicolumn{1}{c|}{-0.266} & -1.305 & 1.000 & \multicolumn{1}{c|}{-0.110} & 2.529 & 0.107 & 0.174 \\
-0.6 & -1.969 & 0.358 & \multicolumn{1}{c|}{-0.211} & -3.202 & 0.022 * & \multicolumn{1}{c|}{-0.386} & -2.520 & 0.109 & \multicolumn{1}{c|}{-0.321} & -2.353 & 0.159 & \multicolumn{1}{c|}{-0.172} & -1.237 & 1.000 & \multicolumn{1}{c|}{-0.096} & 1.261 & 1.000 & 0.086 \\
-0.4 & -1.869 & 0.438 & \multicolumn{1}{c|}{-0.225} & -2.853 & 0.050 & \multicolumn{1}{c|}{-0.333} & -2.270 & 0.190 & \multicolumn{1}{c|}{-0.284} & -1.484 & 0.900 & \multicolumn{1}{c|}{-0.099} & -0.367 & 1.000 & \multicolumn{1}{c|}{-0.041} & 0.757 & 1.000 & 0.064 \\
-0.2 & -1.519 & 0.845 & \multicolumn{1}{c|}{-0.161} & -2.035 & 0.313 & \multicolumn{1}{c|}{-0.223} & -1.235 & 1.000 & \multicolumn{1}{c|}{-0.159} & -0.709 & 1.000 & \multicolumn{1}{c|}{-0.053} & 0.161 & 1.000 & \multicolumn{1}{c|}{0.020} & 0.934 & 1.000 & 0.080 \\
0.0 & -0.867 & 1.000 & \multicolumn{1}{c|}{-0.084} & -0.794 & 1.000 & \multicolumn{1}{c|}{-0.087} & 0.099 & 1.000 & \multicolumn{1}{c|}{0.014} & -0.014 & 1.000 & \multicolumn{1}{c|}{-0.001} & 0.835 & 1.000 & \multicolumn{1}{c|}{0.101} & 1.272 & 1.000 & 0.105 \\
+0.2 & -0.031 & 1.000 & \multicolumn{1}{c|}{-0.003} & 0.302 & 1.000 & \multicolumn{1}{c|}{0.033} & 1.411 & 1.000 & \multicolumn{1}{c|}{0.206} & 0.457 & 1.000 & \multicolumn{1}{c|}{0.034} & 1.618 & 0.707 & \multicolumn{1}{c|}{0.195} & 1.685 & 0.624 & 0.159 \\
+0.4 & -0.185 & 1.000 & \multicolumn{1}{c|}{-0.019} & 0.862 & 1.000 & \multicolumn{1}{c|}{0.098} & 2.051 & 0.303 & \multicolumn{1}{c|}{0.321} & 1.643 & 0.674 & \multicolumn{1}{c|}{0.110} & 2.692 & 0.074 & \multicolumn{1}{c|}{0.319} & 1.935 & 0.384 & 0.200 \\
+0.6 & 0.592 & 1.000 & \multicolumn{1}{c|}{0.062} & 1.525 & 0.835 & \multicolumn{1}{c|}{0.165} & 2.711 & 0.070 & \multicolumn{1}{c|}{0.449} & 1.615 & 0.711 & \multicolumn{1}{c|}{0.099} & 3.195 & 0.022 * & \multicolumn{1}{c|}{0.366} & 2.390 & 0.146 & 0.257 \\
+0.8 & -0.171 & 1.000 & \multicolumn{1}{c|}{-0.017} & 1.865 & 0.441 & \multicolumn{1}{c|}{0.197} & 2.929 & 0.042 * & \multicolumn{1}{c|}{0.455} & 3.425 & 0.012 * & \multicolumn{1}{c|}{0.205} & 3.984 & 0.003 ** & \multicolumn{1}{c|}{0.451} & 2.358 & 0.157 & 0.240 \\
+1.0 & 0.616 & 1.000 & \multicolumn{1}{c|}{0.062} & 2.231 & 0.207 & \multicolumn{1}{c|}{0.217} & 3.314 & 0.016 * & \multicolumn{1}{c|}{0.462} & 2.912 & 0.044 * & \multicolumn{1}{c|}{0.155} & 4.050 & 0.002 ** & \multicolumn{1}{c|}{0.395} & 2.813 & 0.055 & 0.228 \\
+1.2 & 1.428 & 0.991 & \multicolumn{1}{c|}{0.135} & 2.881 & 0.047 * & \multicolumn{1}{c|}{0.247} & 4.038 & 0.003 ** & \multicolumn{1}{c|}{0.495} & 1.705 & 0.601 & \multicolumn{1}{c|}{0.110} & 3.603 & 0.008 ** & \multicolumn{1}{c|}{0.352} & 3.522 & 0.010 ** & 0.241 \\
+1.4 & 1.131 & 1.000 & \multicolumn{1}{c|}{0.116} & 2.950 & 0.040 * & \multicolumn{1}{c|}{0.264} & 4.664 & 0.001 *** & \multicolumn{1}{c|}{0.568} & 2.208 & 0.217 & \multicolumn{1}{c|}{0.150} & 4.500 & 0.001 *** & \multicolumn{1}{c|}{0.456} & 4.437 & 0.001 ** & 0.307 \\
+1.6 & 1.849 & 0.455 & \multicolumn{1}{c|}{0.187} & 1.930 & 0.387 & \multicolumn{1}{c|}{0.191} & 4.078 & 0.002 ** & \multicolumn{1}{c|}{0.527} & 0.130 & 1.000 & \multicolumn{1}{c|}{0.011} & 2.888 & 0.046 * & \multicolumn{1}{c|}{0.352} & 3.468 & 0.011 * & 0.332 \\
+1.8 & 1.400 & 1.000 & \multicolumn{1}{c|}{0.126} & 1.980 & 0.350 & \multicolumn{1}{c|}{0.226} & 2.714 & 0.070 & \multicolumn{1}{c|}{0.377} & 1.111 & 1.000 & \multicolumn{1}{c|}{0.099} & 1.915 & 0.399 & \multicolumn{1}{c|}{0.251} & 1.308 & 1.000 & 0.153 \\
+2.0 & 1.556 & 0.791 & \multicolumn{1}{c|}{0.188} & 1.814 & 0.488 & \multicolumn{1}{c|}{0.240} & 2.463 & 0.124 & \multicolumn{1}{c|}{0.427} & 0.501 & 1.000 & \multicolumn{1}{c|}{0.038} & 1.290 & 1.000 & \multicolumn{1}{c|}{0.213} & 1.247 & 1.000 & 0.184 \\
+2.2 & 2.435 & 0.134 & \multicolumn{1}{c|}{0.393} & 1.743 & 0.561 & \multicolumn{1}{c|}{0.268} & 1.892 & 0.421 & \multicolumn{1}{c|}{0.413} & -1.481 & 0.907 & \multicolumn{1}{c|}{-0.122} & 0.007 & 1.000 & \multicolumn{1}{c|}{0.001} & 0.822 & 1.000 & 0.129 \\
+2.4 & 2.894 & 0.051 & \multicolumn{1}{c|}{0.340} & 2.161 & 0.251 & \multicolumn{1}{c|}{0.257} & 2.718 & 0.075 & \multicolumn{1}{c|}{0.442} & -0.966 & 1.000 & \multicolumn{1}{c|}{-0.098} & 0.456 & 1.000 & \multicolumn{1}{c|}{0.074} & 1.495 & 0.894 & 0.185 \\
+2.6 & 2.864 & 0.056 & \multicolumn{1}{c|}{0.362} & 1.924 & 0.408 & \multicolumn{1}{c|}{0.271} & 3.515 & 0.012 * & \multicolumn{1}{c|}{0.512} & -0.738 & 1.000 & \multicolumn{1}{c|}{-0.089} & 0.770 & 1.000 & \multicolumn{1}{c|}{0.127} & 1.916 & 0.414 & 0.221 \\
+2.8 & 3.545 & 0.012 ** & \multicolumn{1}{c|}{0.443} & 1.919 & 0.416 & \multicolumn{1}{c|}{0.274} & 2.513 & 0.124 & \multicolumn{1}{c|}{0.441} & -1.261 & 1.000 & \multicolumn{1}{c|}{-0.186} & -0.218 & 1.000 & \multicolumn{1}{c|}{-0.035} & 1.548 & 0.823 & 0.165 \\
+3.0 & 4.953 & 0.001 *** & \multicolumn{1}{c|}{0.690} & 2.635 & 0.101 & \multicolumn{1}{c|}{0.383} & 3.545 & 0.014 * & \multicolumn{1}{c|}{0.556} & -1.591 & 0.775 & \multicolumn{1}{c|}{-0.248} & -0.669 & 1.000 & \multicolumn{1}{c|}{-0.110} & 1.307 & 1.000 & 0.140 \\ \bottomrule
\end{tabular}
\end{table*}

The mean and Std. of walking speeds for each scenario are shown in Table~\ref{Tab:mean_std_speed}.
The bold and underlined values indicate the maximum and minimum values of mean and Std. of each time point, respectively.
The mean of walking speeds was highest in \textit{AV w/ eHMI} and lowest in \textit{MV} before the car stopped.
After the car stopped, the participants had the lowest mean of walking speeds in \textit{AV w/ eHMI after PI}.
In contrast, the participants had the highest mean of walking speeds in \textit{MV} after the car stopped for one second. Furthermore, the Std. of walking speeds was the lowest in \textit{AV w/ eHMI after PI} throughout the crossing, \ie from -1.0~[s] to 1.0~[s] and from 2.0~[s] to 3.0~[s]. 
Moreover, the highest Std. of walking speeds was recorded from -1.0~[s] to 1.2~[s] in \textit{MV} and from 1.2~[s] to 1.8~[s] in \textit{AV w/ eHMI after PI} as well as from 2.0~[s] to 3.0~[s] in \textit{AV w/o eHMI}.

This study assumed that the time points in the crossing process and scenarios effected walking speeds.
Therefore, a repeated measured two-way ANOVA was used to estimate the changes in walking speed based on the two variables and their interaction.
The results of the repeated measured two-way ANOVA are shown in Table~\ref{Tab:2way_ANOVA}.
The Greenhouse-Geisser corrected p-values (p-GG-corr) for time points and scenarios as well as their interaction turned out to be smaller than 0.001, 0.01 and 0.05, respectively, implying that the means of time points and scenarios as well as their interaction possessed a statistically significant effect on walking speeds.

On one hand, although Table~\ref{Tab:2way_ANOVA} shows significant differences among time points, considering that the change in walking speed of pedestrians during the experiment is an obvious result, we did not conduct post-hoc pairwise comparisons for walking speed between pairs of time points.

On the other hand, Table~\ref{Tab:2way_ANOVA} also shows significant differences among four scenarios. 
Therefore, we used paired t-tests (two-sided) to compare the differences in the mean of walking speed from -1 to +3 seconds between each pair of scenarios.
These results are shown in Table~\ref{Tab:post-hoc_eHMI}.
We found that during the road crossing process, the walking speed in the \textit{AV w/ eHMI after PI} scenario was significantly lower than in the AV w/ eHMI scenario. In all other paired scenarios, the differences in walking speed were not significant.

To further investigate the simple main effects of four scenarios within time points on walking speeds, a repeated measured t-test was performed as pairwise comparisons for each time point of walking speeds referring to the test approach in~\citet{dey2021communicating}. 
Here, the Bonferroni correction was used to adjust the p-values. 
The results of the post-hoc pairwise comparisons are shown in Table~\ref{Tab:post-hoc}.

\section{DISCUSSION}

\subsection{Subjective feelings of participants}
\label{sec:Dis_Q}

This subsection focuses on the subjective feelings of pedestrians in different four scenarios. 
Firstly, the issues of \textit{AV w/o eHMI} and the effectiveness of eHMI are discussed.
Then, the importance of eHMI pre-instruction is discussed to verify \textbf{H1}.

\subsubsection{Issues of AV w/o eHMI on pedestrians' subjective feelings}

The participants who encountered \textit{AV w/o eHMI} after encountering \textit{MV}, yielded worsened subjective evaluations from Q1-Q6 as shown in Fig.~\ref{fig:result_01} and Table~\ref{tab:QA_Posthoc}.
Specifically, the results for Q1 showed that it was significantly harder to understand the driving intention when encountering \textit{AV w/o eHMI} as opposed to encountering \textit{MV}.
Similarly, Q2 results demonstrate that the driving behaviors of \textit{AV w/o eHMI} became significantly more difficult to predict than the driving behavior of \textit{MV}.
The following reasons could be considered:
(1) In the case of the MV, the dummy driver used the gesture to convey explicit information about driving intentions to the participants after the car stopped.
(2) In the case of \textit{AV w/o eHMI}, understanding the driving intention of \textit{AV w/o eHMI} and predicting the driving behaviors became more difficult for the participants because they could not obtain explicit information from the AV.

In addition, the results from Q3-Q5 indicated that the sense of danger, the trust in car and the sense of relief for \textit{AV w/o eHMI} were significantly worse compared to those of \textit{MV}. 
These findings are consistent with the results from the causal discovery based on the cognitive-decision-behavior model (see Fig.~\ref{fig:model}) and can be explained through the findings in \citep{liu_IV24}, \ie when pedestrians have poor situation awareness of AVs while crossing the road, they are more likely to experience feelings of danger, reduced trust in AVs, and a diminished sense of relief.
Therefore, compared to the \textit{MV}, participants were more cautious when the driving behaviors of the \textit{AV w/o eHMI} were difficult to predict, which in turn increased their subjective risks, such as the sense of danger.
Therefore, the participants demonstrated less trust in the AV's ability to ensure their safety.
Additionally, they found it difficult to cross with a high sense of relief while their sense of danger increased and their trust in \textit{AV w/o eHMI} decreased.

Finally, based on the results of Q6, the deterioration of situation awareness and subjective feelings, \eg sense of danger and trust in the car, may have reduced participants' ability to quickly make decisions, leading to increased hesitation when crossing in front of \textit{AV w/o eHMI}.
This finding also aligns with the causal discovery results in~\citep{liu_IV24}.

Overall, these research findings suggest that there may be concerning and thought-provoking issues when introducing AVs without eHMI into society.

\subsubsection{Effectiveness of eHMI in improving the subjective feelings of pedestrians}

As shown in Fig.~\ref{fig:result_01} and Table~\ref{tab:QA_Posthoc}, participants, who encountered \textit{AV w/ eHMI}, were significantly able to understand the driving intention better and to predict the driving behavior than when they encounter \textit{AV w/o eHMI}.
From the results of Q1 and Q2, this study suggests that the participants could easily understand the driving intentions of the AV through the eHMI and predict the driving behaviors of the AV. 

Furthermore, this study confirmed that using eHMI, \ie \textit{AV w/ eHMI}, could significantly improve participants' sense of danger, trust in AV, and the sense of relief, compared to \textit{AV w/o eHMI}.
The above results are consistent with~\citep{faas2020external,liu_IV24}, which reported that the eHMI contributes to more positive feelings such as a sense of safety and trust in AV.

In addition, those results could be due to the current driving state and the driving intentions in the future of the AV which were clearly informed through the eHMI. In particular, the clear text message ``UGOKIMASEN'' read as ``AV does not move'' on the eHMI which was designed to be displayed after the car stopped.
Therefore, participants received consistent implicit information about the car's behavior and explicit information from the eHMI.
As a result, their sense of danger decreased and trust in AV as well as the sense of relief improved during their crossing. 

Furthermore, these enhancements in subjective feelings facilitated participants in making decisions with significantly reduced hesitancy in the \textit{AV w/ eHMI} compared to \textit{AV w/o eHMI}. 
These results suggest that eHMI is an important and effective communication method for pedestrian--AV interaction. 

In addition, Fig.~\ref{fig:result_01} and Table~\ref{tab:QA_Posthoc} show that the evaluation results of Q4 and Q5 for \textit{AV w/ eHMI} were still significantly lower than their evaluation results for \textit{MV}.
This indicated that although the eHMI improved participants' trust in AV and sense of relief during crossing when they encountered the \textit{AV w/ eHMI}, it did not reach the subjective feelings as when they encountered \textit{MV}.

Moreover, when comparing the results of Q1, Q2 and Q6 for \textit{MV} and \textit{AV w/ eHMI}, we found there were not significant differences between these two scenarios.
This indicates that the application of eHMI in AV had a similar effect to the communication gestures of human drivers. 
It could assist pedestrians in situation awareness and quickly making decisions when crossing the road.

\subsubsection{Importance of eHMI pre-instruction for improving the subjective feelings of pedestrians}

When encountering \textit{AV w/ eHMI after PI}, the participants were more easy to understand the driving intentions and to predict the driving behaviors than before the pre-instruction, \ie \textit{AV w/ eHMI} (see  Fig.~\ref{fig:result_01}).
These results also prompted participants to perceive significantly less danger from the driving behavior and feel a significantly greater sense of relief regarding \textit{AV w/ eHMI after PI}, compared to the conditions before the pre-instruction, \ie \textit{AV w/ eHMI}.
However, we observed no significant differences in the outcomes of trust in AV (Q4) and the hesitancy in decision making (Q6) before and after the pre-instruction.

From another perspective, to compare the \textit{MV} with the \textit{AV w/ eHMI after PI}, there were no significant differences for results of Q1 to Q6. 
This indicates that when employing eHMI on the AV and providing pedestrians with pre-instructions, their situation awareness, subjective feelings, and decision-making hesitations could be enhanced to match the criteria observed in interactions with the MV.

Referring to the cognition-decision-behavior model (see Fig.~\ref{fig:model}) proposed in~\citep{Liu2022_APMV}, the above findings provide robust validation for \textbf{H1}.
Specifically, pedestrians who correctly understand the rationale of eHMI through the pre-instruction exhibit improved situation awareness, subjective feelings, and decision-making during the interaction. 
Furthermore, these results clearly indicated the effectiveness and importance of using eHMI and the pre-instruction to improve the subjective feelings of pedestrians who encountered the AV.

\subsection{Walking behaviors of participants}

Firstly, Table~\ref{Tab:2way_ANOVA} shows that the learning effect did not affect the average walking speed of pedestrians while crossing the road under the four scenarios because there was no significant difference in the average walking speeds under the four scenarios, except between \textit{AV w/ eHMI} and \textit{AV w/ eHMI after PI}.
Meanwhile, Table~\ref{Tab:2way_ANOVA} also indicates that the fatigue effect, which could arise from repeated experiments, did not affect the average walking speeds. 
Specifically, there is no significant difference in average walking speed between the first experienced scenario, \ie \textit{MV}, and the last experienced scenario, \ie \textit{eHMI w/ eHMI after PI}.
After confirming that walking behaviors were not significantly affected by the fatigue effect and learning effect, the following subsections discuss the communication and guidance effects of eHMI pre-instruction on walking behaviors in the four scenarios to verify the \textbf{H2}.

\subsubsection{Communication effect of the eHMI pre-instruction on walking speed}
Table~\ref{Tab:mean_std_speed} shows that the mean of the walking speeds of the participants in \textit{MV} was the lowest before the car stops. Furthermore, the figure indicates that it was the highest for most of the time after the car stopped at 0.6~[s] to 3.0~[s]. 
Specifically, the mean of the walking speeds in \textit{MV} was significantly higher than that in \textit{AV w/ eHMI after PI} from 0.8~[s] to 1.6~[s] (see Table~\ref{Tab:post-hoc}). Two reasons were considered for the above differences. 
First, the interpretation could be based on Japanese culture, given that all the participants are Japanese and do not like to burden other people, in general. 
Therefore, in the case of manned MV, the participants likely crossed quickly after the dummy driver presented "After you" using the hand gesture to reduce the burden for the dummy driver. 
Conversely, it can be inferred that participants crossed slowly because they didn't have to consider the burden they might impose on others, because the AV was a driverless car.
Furthermore, the participants crossed the road slowly \textit{AV w/ eHMI after PI} with a higher sense of relief (see Fig.~\ref{fig:result_01}) because they fully understood that the AV did not move by displaying the message ``UGOKIMASEN'' on the eHMI. 

Second, the difference in the meanings of messages sent by the dummy driver and the eHMI. 
Referring to the study~\citet{li2021autonomous}, the message ``After you'' could be clarified using the hand gesture of the dummy driver. 
Particularly, this suggests that the gesture could facilitate the pedestrians' decision-making. 
On the other hand, the message ``UGOKIMASEN'' indicated the state of the AV (\ie ``AV does not move now''), which can inform the pedestrians and help them understand the situation. 
Moreover, this study observed that the suggestion provided by the dummy driver was relatively urgent, while the information provided by the AV was not as urgent as it allowed the participants to make their own decisions.

In addition, this study reveals that the mean of walking speeds in \textit{AV w/o eHMI} was significantly higher than that in \textit{AV w/ eHMI} and \textit{AV w/ eHMI after PI} from 0.6~[s] to 1.6~[s] as shown in Table~\ref{Tab:post-hoc}. Referring to the results of the subjective evaluation that are illustrated in Fig.~\ref{fig:result_01}, this study considered that participants felt relief and trusted the AV when they could understand the correct driving intentions of the AV through the eHMI after the pre-instruction.
Therefore, they could pass the road in a slow and relaxed manner,
which is in contrast to the result in~\citet{kooijman2019ehmis};
they reported that the mean of pedestrians' walking speeds was faster than that for \textit{AV w/o eHMI} after the AV slowed down while using a text eHMI to indicate yielding. 
We consider the difference between \citet{kooijman2019ehmis} and this study lies in the display time of the eHMI; while \citet{kooijman2019ehmis} displayed the eHMI while decelerating, this study displayed eHMI after the car stopped.
The present experiment found that the participants felt safe after displaying eHMI, while in \citet{kooijman2019ehmis}, the participants may still perceive risk even after the display as the AV continues to approach. This study considered that this difference in perception of risk might cause a difference in walking speed.

\subsubsection{Guidance effect of the eHMI pre-instruction on the variation of walking speed}

To verify the hypothesis \textbf{~H2}, we focused on comparing the variation of pedestrians' walking speeds before and after the car stopping in different scenarios. 
As Table~\ref{Tab:mean_std_speed} demonstrates, this study confirms that the Std. of participants' walking speeds in \textit{AV w/ eHMI after PI} was lowest within 1 second before and after the car stopped, respectively, from -1.0~[s] to 1.0~[s] and 2.2~[s] to 3.0~[s].

After the AV stopped, (\ie after the eHMI appeared), the Std. of the participants' walking speeds within one second was lowest. Particularly, this indicates the effects of information guidance on using eHMI which were based on the participants' walking speeds exhibiting a small variation over the time horizon in multiple trials when they fully understood the principle of eHMI from the pre-instruction. 
It can also be considered that, as shown in Fig.\ref{fig:model}, once pedestrians formed a correct mental model of interaction with AVs, their decisions became consistent, resulting in similar interaction behaviors each time.

In summary, the hypothesis \textbf{~H2} has been confirmed, indicating that participants' walking behavior becomes more consistent over multiple interactions as they gain a comprehensive understanding of eHMI principles through the pre-instruction.

\subsection{Outlook}

Through a user experiment, this paper validated that pre-instruction of eHMIs was crucial in the negotiation between pedestrians and an AV.
Given that there is currently no unified design standard for eHMIs in the industry, we hope that in the near future, various car manufacturers can provide pre-instruction on eHMIs to various traffic participants through television commercials.
For example, a television commercial could showcase demonstrations of AV-pedestrian interactions and explain their interaction logic, including the sensor detection range, the information displayed by eHMI in different scenarios, and the meaning conveyed by this information.
We believe that this is the most feasible and practical method to achieve a large-scale pre-instruction for eHMIs.

Additionally, various studies are currently focusing on using machine learning-based methods to model interactions between pedestrians and AVs to improve pedestrian behavior prediction~\citep{Saleh_2018, Yang_2022, Zhu_2023}. 
However, current autonomous driving systems are not fully reliable in pedestrian detection and behavior prediction~\citep{Zhang_2023, Cummings_2024}.
We consider that the diversity of decisions and behaviors made by pedestrians during interactions is one of the reasons contributing to this issue.
To address this, we believe that providing pedestrians with the AV's driving intentions via eHMI can reduce the diversity of their decisions and behaviors. 
Specifically, this paper validates that pre-instruction can better facilitate eHMIs in guiding pedestrian walking behavior.
Therefore, looking ahead to the future, using eHMI in interactions between pedestrians and AVs may indirectly improve the accuracy of AV in predicting pedestrian behavior, as also mentioned in~\citep{li2021autonomous}.

\subsection{Limitations}

Although this study focused on verifying the influence of eHMI pre-instruction on pedestrians' subjective perceptions and walking behavior, the use of other types of eHMI, such as icons and colored light bars, has the potential to provide participants with a different interaction experience than the text-based eHMI used in this experiment.

Using a within-subject experimental design helps evaluate the impact of eHMI pre-instructions on participants' psychological and behavioral responses when interacting AVs.
However, as reasons discussed in Section \ref{sec:order}, each participant experiences a fixed sequence of scenarios, which may introduce potential order effects on the results.

Furthermore, cultural influences on the experimental results cannot be disregarded, as all participants were Japanese.
Although the participants' age range is from 23 to 68 years old, the results may not generalize to other age groups, such as children or people over 70 years old.

\section{CONCLUSION}

This study investigated the effects of pre-instruction of eHMI on subjective feelings and walking behaviors of pedestrians and designed an interaction experiment in a road crossing scenario.

A clear issue was found when participants encountered \textit{AV w/o eHMI}, they had difficulty understanding the driving intention and predicting the behavior of the AV after becoming accustomed to \textit{MV}. This led to worsened subjective feelings, such as increased danger, reduced trust, and less relief, which caused hesitancy in decision-making when crossing the road.

In addition, the use of eHMI on the AV helped pedestrians better understand its driving intentions and predict its behavior, improving their subjective feelings and decision-making hesitation. However, despite these improvements, participants' evaluations of \textit{AV w/ eHMI} did not surpass those of \textit{MV}.

Pre-instruction on the rationale behind eHMI helped participants understand the AV’s intentions and predict its behavior, improving subjective feelings and decision-making, similar to interactions with \textit{MV}. 
Additionally, the eHMI reduced variation in participants' walking speeds across trials once they fully understood its principles through the pre-instruction.

In future, the negative effects of eHMI on pedestrians should be studied, such as overtrust~\citep{hollander2019overtrust,kaleefathullah2020external} and distraction~\citep{lee2021negative}. 
Particularly, these issues will be solved by refining the pre-instruction and analyzing its influence on pedestrians, including providing the pre-instruction for cognition and risk evaluation. 
Finally, the standardization of eHMI should be realized to ensure the design of a more accurate method of pre-instruction.

\section*{ACKNOWLEDGMENT}

This work was received funding from the Japan Society for the Promotion of Science (JSPS) KAKENHI Grant Numbers 20K19846 and 19K12080, Japan; and partly supported by Toyota Motor Corporation, Japan.
The authors express their appreciation for the valuable advice provided by Mr. Masaya Watanabe from Toyota Motor Corporation.

\section*{CRediT authorship contribution statement}
\textbf{Hailong Liu}: Conceptualization, Investigation, Resources, Methodology, Validation, Formal analysis, Visualization, Writing - Original Draft, Writing - Review \& Editing, Project administration.

\textbf{Takatsugu Hirayama}: Conceptualization, Methodology, Validation, Writing - review \& editing.

\bibliographystyle{IEEEtranN} 
\footnotesize
\bibliography{sample-base}

\begin{IEEEbiography}
 [{\includegraphics[width=1in,height=1.25in,clip,keepaspectratio]{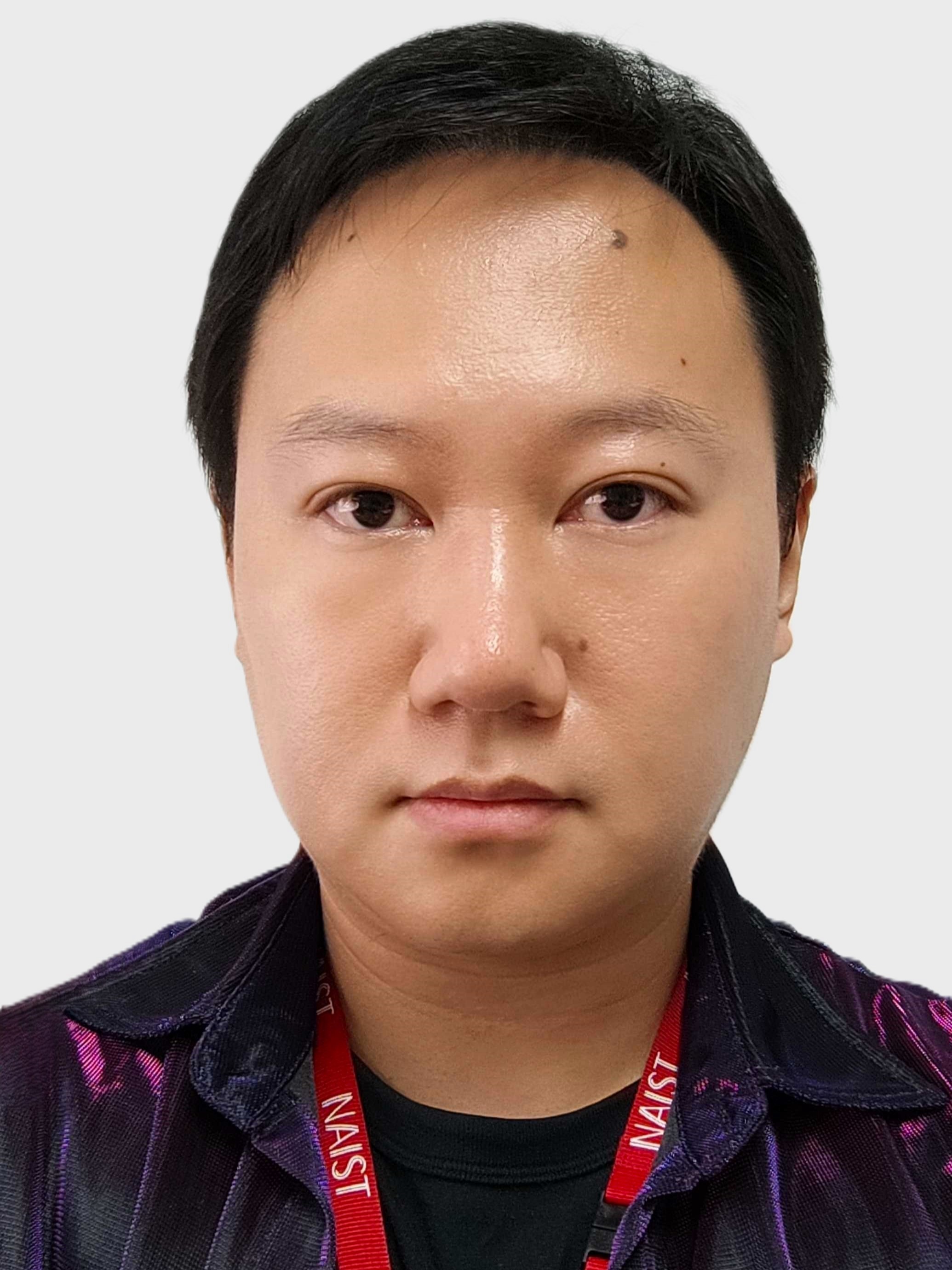}}]{Hailong~LIU}  (Member, IEEE)
 received his M.Eng. and Ph.D. degrees in Engineering from the Graduate School of Information Science and Engineering, Ritsumeikan University, Japan in 2015 and 2018, respectively. From April 2016 to March 2018, he was a JSPS Research Fellow for Young Scientists (DC2). He worked as a researcher at Nagoya University, Japan, from April 2018 to October 2021. From November 2021 to January 2024, he was an Assistant Professor at the Graduate School of Science and Technology, NAIST, Japan, and was promoted to Associate Professor in February 2024.
 His research has focused on human factors and machine learning in the intelligent transportation systems. 
 He was awarded the IEEE IV Best Student Paper Award in 2015, and the IEEE GCCE Outstanding Paper Award in 2016.
 He is a member of IEEE, JSAE, JSAI, and SICE. 
 He is also a committee member of IEEE ITSS TC on Human Factors in Intelligent Transportation Systems.
\end{IEEEbiography}

\begin{IEEEbiography}
 [{\includegraphics[width=1in,height=1.25in,clip,keepaspectratio]{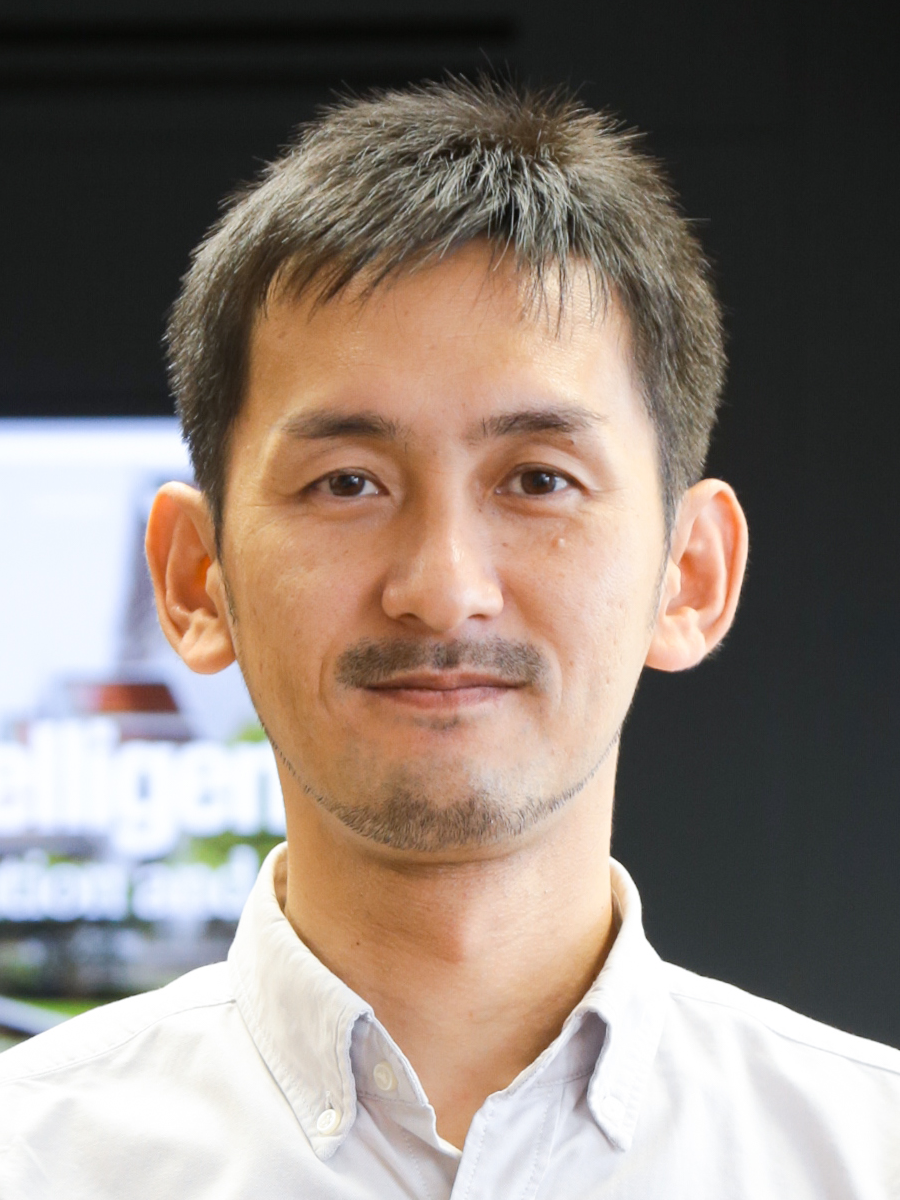}}]{Takatsugu~Hirayama} (Member, IEEE)
is a Professor at University of Human Environments and a Visiting Professor at Graduate School of Informatics, Nagoya University, Japan.
He received his M.Eng. and Ph.D. degrees in Engineering Science from Osaka University in 2002 and 2005, respectively. From 2005 to 2011, he was a Research Assistant Professor at the Graduate School of Informatics, Kyoto University. In 2011, he moved to the Graduate School of Information Science, Nagoya University. He was an Assistant Professor from 2012 to 2014, and a Designated Associate Professor from 2014 to 2017. From 2018 to 2021, he was a Designated Associate Professor at the Institutes of Innovation for Future Society, Nagoya University. His research interests include computer vision and human-computer interaction. He is a member of IEEE, ACM, IEICE, and IPSJ.
\end{IEEEbiography}

\end{document}